# A wireless hand-held platform for robotic behavior control

*Christopher A. Tucker*

cartheur@gmail.com

Abstract

The need for customizable properties in autonomous robotic platforms, such as in-home nursing care for the elderly and parallel implementations of human-to-machine control interfaces creates an opportunity to introduce methods deploying commonly available mobile devices running robotic command applications in managed code. This paper will discuss a human-to-machine interface and demonstrate a prototype consisting of a mobile device running a configurable application communicating with a mobile robot using a managed, type-safe language, C#.NET, over Bluetooth.

# Introduction

The advancement of autonomous robotic technology, which allows the ordinary user to control a machine's actions, as well as contributing to its programming, presents challenges based on environment, scenario, and desired human-to-machine interaction. For the purposes of this paper, one specific example is the prevalence of in-home companion robots. While the form might very well be that of a wheeled machine with a friendly electric face and function to perform tasks for those not entirely able to take care of themselves, such as those in care-giver environments, the culmination of said requires thought and implementation to anticipate diversity. While the most obvious control of such a machine would be voice, neglects aspects of the scenario wherein such an interface would fail or would require alteration to suit the peculiar needs of its deployment. Where tightly integrated interaction is imperative, the interface described in this paper bridges the gap between voice and tactile control normally affixed to a robot.

An example in-home companion robot is one created during the project *CompanionAble* [1]. This machine is a wheeled device possessing what resemble electronic eyes responding to vocal commands and through a touch-screen interface on its chest, when folded outward, looks like a duck. However, what is not immediately obvious are the alternatives to communicating with the machine for those persons unable to be heard clearly by the robot in the form of commands nor able to go to the robot to input commands directly via the touch screen. Additionally, what other designs are possible wherein a user can communicate with the robot? Is it possible to pre-design some sort of program that could anticipate a collection of actions described in possible scenarios in the environment? The answer is certainly, yes. If so, then, what kind of form would such a piece of technology take? In this paper, one such in particular is proposed of a commonly available mobile device such as a smartphone over Bluetooth radio paired to a robot, whose form is a fixed crane on a rotating pad with claw, controlling its actions at a distance via clicking buttons within application software running on the handset.

Finally, one last concern is the programming language in which the application is written. As technology such as the robot listed above would be deployed in a server environment where medical monitoring over web-services takes place, such systems are increasingly becoming conformant to Microsoft-based software and operating systems. For maximum compatibility, the application described in this paper is written in C#.NET and is fully configurable and expandable. One such immediate addition would be panic alarming for alert emergencies and the inclusion of location-based services so medical help could locate the individual in need via the handset or the robot tasked to protect them.

The main idea of this proposal is to present a programming context and structure such that a widely adopted and easily available platform for program extensions by sophisticated users, who do not necessarily need to be familiar with the complicated aspects of robot behavior. The goal is programs and autonomy can be adjusted and complimented to aid in the adaptation of the robot to its environment and desired interaction with humans to a greater or a less degree.

## The human-machine interface

The need for a machine caregiver presents challenges to both form and function. In consideration of humans, the form of the robot follows its intended purpose; function consist the necessary grand and subtle details to satisfy the necessity of purpose. Both the appearance and projected emotion inspired upon its viewing and the expected behaviors are the most important to consider rather than the physical construction of the machine in all its intellectual wonderment.

Briefly then, what is the ideal form and function of an in-home care-giving robot? A form such as *CompanionAble* has ease of utility by its package that is wheels instead of legs, and presents behaviors to the user as a pre-programmed set of responses to anticipated scenarios inherent in the environment. Whether or not such a form is superior to other types of robots is outside the scope of this paper, and will not make arguments save to note that a fold-out touch-screen interface to send commands is less than ideal anticipating the users in the environment will most likely be immobile than otherwise and may not possess the ability to properly execute commands. Rather, this paper will detail an added dimension of function to the existing form by the introduction of a remote keypad deployed on a mobile device hosting Windows Phone.

Targeting the platform in such a way and the particular choice reflects the prevalence of enterprise systems designed to assist in duty-of-care scenarios where the use of C# manifest as web services and the strength and ease of using the .NET Compact Framework to extend the model into mobility.

There is sufficient need for an alternative to speech-driven commands for in-home companion robots. Typically, accuracy errors and efficiency peaks of 80% are observed whereas humans can discriminate errors between 2 - 4%. Additionally, aging or disabled patients are quite likely unable to speak loudly or plainly enough for the robot to respond to the intended command [2, 3]. Although the problems with voice-recognition software are not directly addressed in this paper, advances in Turing machine technologies could aid in solving these sets of problems [4 - 6]. There is a quality of using a mobile phone for voice recognition in that the hardware package is designed specifically with the human voice in mind. This may or may not be superior to sensitive microphones installed in the robots, however, it nevertheless provides a degree of traceability of hardware components justified for use in the manner that they are. Additionally, a further interface is desirable.

As such, the interface ideally should be a combination of a voice-recognition, language-dependent speech in the form of text-to-speech (TTS), and a remote-tactile interface, which uses individual tasking, such as motor control and sensor feedback, and possessing the ability to be programmed to execute serial tasks based on symbolic representation.

# Managed code, native code, and the compact framework

Managed, type-safe code is desirable when the programming task is to insure maximum reliability, robustness, and repeatability of code execution. Native code, while intrinsically more powerful, if not programmed properly can lead to undesirable results and unwanted behavior not predicted in the ethological model. Since it is expected numerous programmers over lengthy periods would be enhancing, detracting, and modifying the application, establishing rule-sets for the methodology of programming is a sound notion. Not only this, but it is further suggested a type of integrated development environment (IDE) to be created wherein programmers simply wanting to enhance the application by building more sophisticated execution tasks (behaviors) could do so in an easy and straightforward manner.

Microsoft Windows is quite different between its two main flavors Windows Compact Edition (CE) and Win32/64. While newer versions of the product—Embedded, 7—create additional divergence, revolving an application around Windows Mobile 6 in both its Standard and Professional versions, Windows Phone 7 and 8, in tandem with Windows XP, 7, and 8 covers about 95% of what would be expected platform-wise.

While most of the application described in this paper predominantly makes use of the .NET Compact Framework 2.0, it does use P/Invoke to call Win32 functions where necessary. Many features of .NET present in PC versions of Windows are simply missing in its compact version. A mobile phone is in many ways similar to a computer; using ARM architecture there is the possibility of less memory and therefore a smaller programmatic footprint. Nevertheless, the author of this paper decided that, where possible, to implement the code using the Compact Framework and only using P/Invoke where no other solution existed or would cost an ostensive amount of development time. More detail on this topic is given in [7].

*Compact framework quirks*

One of the glaring issues with Compact Framework is the occasional lack of some designer components for the Windows Form (WinForm), the windowing palette of the application. For example, when using the drag-and-drop feature of Visual Studio to place components onto the WinForm, the necessary dependencies are not applied as properties. The problem is when placing the custom components for the Lego NXT brick, the designer ignores the integration of the brick with each of its motors A, B, and C, and with each of its four sensors. In the code-behind of the WinForm, methods can be written to overcome the discrepancy creating the relationships by such methods as *IntializeBrick()* and *InitializeSensors()*, but including every static property can prove tedious. However, once written into the constructor of the application, the connections perform well. Clamping of sensors and *MethodInvoke()* of their delegates in context with the brick in the Compact Framework were not performed for the writing of this paper. However, their integration and data feedback via a primitive type of fusion would be the next logical step of this project. Regardless, sensors and their implementation are outside the scope of this paper.

Despite the conundrum that an application written for a platform such as Windows 7 or 8 is not necessarily conformable to Windows Phone and the challenges of making such a transition while keeping to best practices and code robustness, it was discovered a sufficient result could be obtained by examining the problem purely in terms of the Windows mobile platform. This strategy is applied during the course of this paper when discussing classes and programmatic logic.

There are two predominant aims here: those necessitated by design and those predicated by functional issues. While appearance and ease of interface is preferred for the former, an elegance and smoothness including power efficiency were considered. While the initial inclination was to split the aims into two separate projects, instead, they were combined and presented to emphasize the need for a package mentality. That is, a compromise between form and function given the limitations of the technology.

## Connections over Bluetooth

In the context of the human-to-machine interface problem, it is not immediately obvious the task of using a touch pad to control a robot at a distance should instead be accomplished wirelessly. While a wired connection might be more advantageous in some manner—clean communication channel with near 100% retention, reduction of electromagnetic radiation—the author argues the use of a hard connection only complicates the environment for not only the robot in the context of its programming but for the people participating with it. In the instance of *CompanionAble*, the foldout screen and the suggestion of its detachable feature is such a hard connection. Rather, a purely wireless solution should be deployed of a one-to-one relationship between the intentions of the user, i.e., what kind of care they require, and the actions of the robot. In human-to-human terms, the requirement is answered by a response of care to alleviate the need, the precursor of the requirement. In such cases, a priori knowledge of how to structure the response is commonplace. To reduce the magnitude of error in giving a skewed or poorly structured response, there are two options: 1. Allow the user to key a device near to his or her proximity to initiate the intended response, 2. Instigate a serial response via a symbolic interpreter. The first is rather straightforward and demonstrated later in the paper in the prototype, the second requires the storage of medical routines programmed into the software to satisfy the a priori knowledge requirement to maximize the success of correctly structured responses.

For the sake of simplicity, a model of a wireless solution over Bluetooth radio was deployed. Bluetooth is an excellent medium to transmit via radio waves instruction sets from an application to a remote listening device. The range is typically 10 meters but it can also transmit though walls found in typical homes, which adds an extra dimension to its utility.

In the context of the prototype, both the mobile phone and the Lego NXT has Bluetooth version 2. Once pairing is established between the devices, that is, from phone to NXT using a passkey, a bidirectional connection can be initiated by creating an instance of Compact Framework's SerialPort class to communicate over a COM port on the mobile device. Once complete, the NXT will listen to commands sent to it from the application.

The NXT, according to documentation of its software development kit (SDK), uses specific types of code to tell the ARM microprocessor what tasks to execute to what components [8]. The brick, shown in Fig. 1, is the "brain" of the robot in that it contains the ARM, memory, Bluetooth, controllers, and operating system.

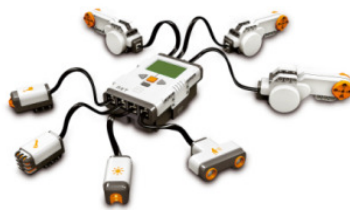

**Fig. 1:** NXT brick

The NXT is a reasonably priced and accessible technology to construct different robotic platforms for scientific investigation. While this paper only deals with one particular form, it does not suggest other alternatives are less important. Rather, it would be interesting to construct numerous forms, classified into types, and see how an application could be used to control the form in the given range.

## Prototypical components

Commonly-available, off-the-shelf technology provides a cost-effective yet powerful platform to solve complex issues with configuring and extending manufactured social robots within human proximity. The Lego Mindstorms NXT is a robotic toy which consists of: a battery-powered brick containing an ARM 32-bit RISC with a USB 2.0 port, a CSR BlueCore 4 version 2 chip—a 16-bit integrated processor runs the Bluetooth stack—and seven outputs, three for servo-controlled motors with angular indicators and four for sensors. These sensors are sound, sonar, color, and bump. In the prototype demonstrated herein, the sensors are not be deployed. By combining the brick with its motor components and modeling them in different configurations, alternate interpretations of locomotive robotic scenarios can be illustrated. For the purposes of this paper, however, the form will be a fixed-arm crane on a rotating pad with a grasping claw using three motors.

Control of the NXT is accomplished via the Bluetooth stack over a mobile phone. Such a means to control a NXT is not new in terms of Java applications, nevertheless, it was more desirable in terms of this project to construct a C# application [7]. The project was manifest in the form of a solution in Visual Studio. The project organization is shown in Fig. 2.

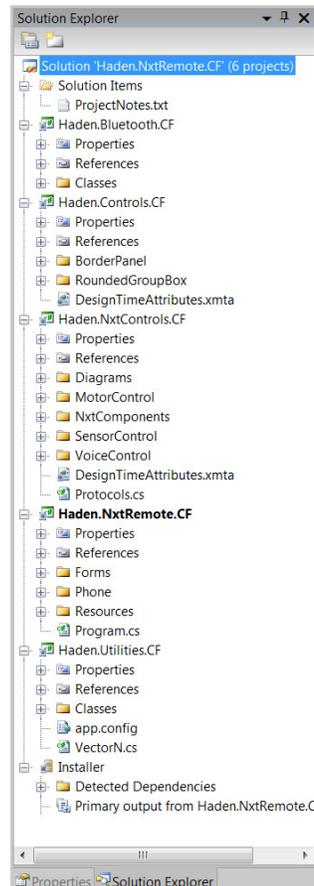

**Fig. 2:** Project Hierarchy

The solution consists of five projects whose main application is the fourth project *Haden.NxtRemote.CF* that contains the windowing elements (in the folder 'Forms') for the desired user experience previously discussed. The other projects are concerned with aspects of the application such as communication, controls, and utilities. The first project from the top, *Haden.Bluetooth.CF*, contains enhanced and data management methods for types of robots other than the NXT. The second project contains custom-controls that give the application a look-and-feel more detailed than that found in default configurations of the Visual Studio toolbox. The third project contains the compendium of dynamic controls used directly in the NxtRemote project such as the motor controls discussed later. In this version, the sensor components are included as well as a primitive form of voice control. The fifth project, *Haden.Utilities.CF*, contains in this version only class and instruction manipulation tasks, however, also contains code for the building of an SDK environment wherein programs can be written and loaded into the application.

The application is hosted on a Windows Phone possessing a touch-screen 240x320, foldout keyboard, landscape view, ARM, Bluetooth, WiFi, and GSM. It has enough RAM to run programs written for the .NET Compact Framework.

The purpose of the program is:

1. To run economically and without disruption to other phone applications a human-to-machine interface over Bluetooth radio,
2. to control a robot at a distance, in this example a single NXT brick,
3. to maintain the connection and keep the components refreshed so that the program can run continuously and without interruption,
4. to demonstrate the utility of such a configuration, its economy, and ability to be adapted to robots of similar form to the ARM architecture.

The operational flow of the program will be constrained to in typical circumstances occurring in the user experience.

*Operation—The user experience*

The robot was designed to be controlled by a single-handed, single-click input on the handset and to run pre-designated tasks in the form of a collection of inputs flowing as a sequence by means of a stored program routine. The use of a stylus to click the buttons is also available. Presently, this version 2.3 of the application includes the following features:

1. Installable *.cab file built under WM 6 Professional SDK and tested in WM 6.1,
2. makes use of Bluetooth pair between device (phone) and robot,
3. auto-connects to the robot on startup,
4. offers manual directional control of three motors,
5. can connect to four sensors in real-time (in development),
6. has command-line interface with the application kernel (in development),
7. is a "friendly" application in that it can be minimized to an icon in the taskbar running in the background,
8. can cleanly disconnect its communication with the robot, and,
9. can be programmed to perform autonomous operations by loading schemas.

The design of the user interface, designed to be simple and functional, is illustrated in Fig. 3:

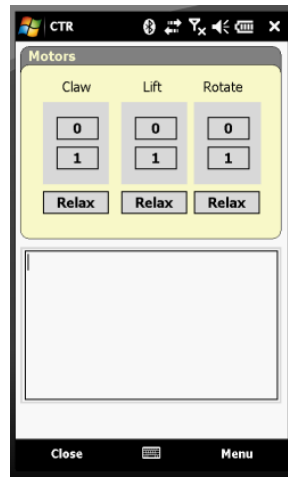

**Fig. 3:** The UI

The group box on the top-half of the form contains controls to three motors labeled to the components used in the robotic model: a claw on a crane with the ability to lift and rotate about a fixed axis. Pressing the '0' key in a block will engage the corresponding motor in one direction while the '1' key in a block will engage it in the opposite direction. The 'Relax' key is designed to put the motor into coasting mode.

The border panel on the lower half of the form contains a command-line text box to send raw commands to the brick. It has no interpreter so can only read and write in hexadecimal.

Application access via a 'Menu' in the bar to the lower right is given to the disconnection of the device with the robot and access to an about screen displaying application and phone details. The screen also contains a further button 'Information'. Minimizing and exiting the application is controlled via access to 'Close' to the lower-left. The application can be minimized while it is connected to the robot, while in this state the runtime is still maintained.

*Logic—The program*

C#.NET is an imperative object-oriented language using a common language interface (CLI) and compiler. Applications written in it operate within a runtime allocated by the operating system and can do so under two main flavors: Win32 and Compact Framework. In its version 2.0, the latter, in having a small footprint is ideal to deploy applications on mobile handsets containing the OS.

A class diagram showing the core constructs of the application is shown in Fig. 4.

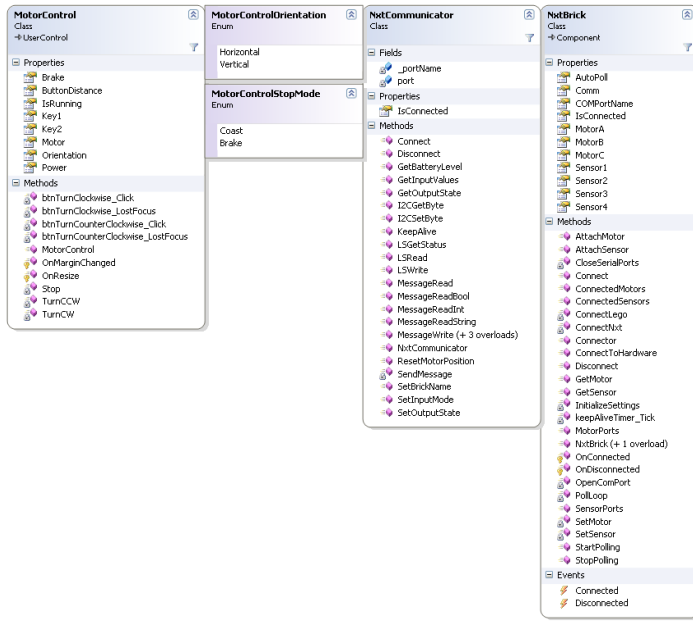

**Fig. 4:** Class Diagram

To the left of Fig. 4 is the motor control. The click events being obvious as they are tied to actions in the buttons shown in Fig. 3, they call the methods *TurnCW()* and *TurnCCW()* respective to the directions they represent. This high-level view of the programming turns otherwise complicated code derived from working with the brick SDK encapsulated in the *Protocols.cs* file.

The NXT is sent commands in the form of messages to Bluetooth mailboxes. These messages can be in the form of text, numbers, Booleans, or byte arrays. An example of the sending of a text message is:

```
public void MessageWrite(byte mailbox, string value) {
            MessageWrite(mailbox, Encoding.ASCII.GetBytes(value));
        }
```

While the sending of a byte array:

```
public void MessageWrite(byte mailbox, byte[] data) {
            if(data.Length > 57) {
                    throw new ArgumentException("Message size must be less than 58 bytes.");
            }
            byte[] message = new byte[5 + data.Length];
            message[0] = 0x80;
            message[1] = (byte)NxtCommand.MessageWrite;
            message[2] = mailbox;
            message[3] = (byte)(data.Length + 1);
            for(int i = 0; i < data.Length; i++) {
                    message[i + 4] = data[i];
            }
            SendMessage(message);
        }
```

The content of these messages, contained in *Protocols.cs*, are in the form of enumerations by type, that is:

```csharp
[Flags()]
public enum NxtMotorMode : byte
{
    None = 0x00,
    MotorOn = 0x01,
    Brake = 0x02,
    Regulated = 0x04
}
```

However, this is not visible to the programmer; rather, it is encapsulated in objects:

```csharp
public void Turn(int speed, int degrees) {
              if(Brick == null) {
                     throw new InvalidOperationException("This motor must be connected to a brick.");
              } else {
                     SetOutputState(
                            speed,
                            NxtMotorMode.MotorOn | NxtMotorMode.Regulated,
                            NxtMotorRegulationMode.MotorSpeed,
                            0,
                            NxtMotorRunState.Running,
                            degrees
                     );
              }
       }
```

As such, by constructing and calling the following methods for turn clockwise:

```csharp
private void TurnCW()
      {
          if (Motor == null)
          {
              MessageBox.Show("There is no motor connected to this control.", "Error", MessageBoxButtons.OK, MessageBoxIcon.Exclamation, MessageBoxDefaultButton.Button1);
          }
          else
          {
              if (!IsRunning)
              {
                  Motor.Turn(-Power, 0);
                  IsRunning = true;
              }
          }
      }
```

and turn counter-clockwise:

```csharp
private void TurnCCW()
      {
          if (Motor == null)
          {
              MessageBox.Show("There is no motor connected to this control.", "Error", MessageBoxButtons.OK, MessageBoxIcon.Exclamation, MessageBoxDefaultButton.Button1);
          }
          else
```

```
        {
            if (!IsRunning)
            {
                Motor.Turn(Power, 0);
                IsRunning = true;
            }
        }
    }
```

and the necessary stop with either a brake or coast feature:

```
private void Stop()
    {
        if (Motor != null && IsRunning)
        {
            if (Brake)
            {
                Motor.Brake();
            }
            else
            {
                Motor.Coast();
            }
            IsRunning = false;
        }
    }
```

yields the functionality necessary to the scope of the application. Because the servomotors have angle and speed indicators, encapsulation of this feature is easy using the appropriate methods and their commands.

On a final note, in order to keep the application always connected to the robot, a keep-alive timer was implemented insuring that the connection will remain until disconnected by a user command.

### Enhancement of the user interface

Since this application was authored for use in a Windows mobile device, it was considered how the application would not only affect the phone, in the manner of power consumption and disk space, but the degree of how it cooperated or not with other applications on the phone. The reason for this was that the application was written for a mobile phone that would most likely have other required applications both in context for the user in what he or she installed but also for the in-home caregiver model. Perhaps, an alarming feature in the form of a panic button is required which is only triggered if running in the foreground.

Therefore, to increase the friendliness of the application co-existing with other applications running on the phone, the application can be minimized to the taskbar. In this state, it appears as a clickable icon. However, the application is still running but in the background. The connection to the robot is maintained and any programs running would continue to until the application is maximized and the appropriate commands given.

# Conclusion and suggestions for future research

This paper discussed a human-to-machine interface and demonstrated a prototype consisting of a mobile device running a configurable application in C# and .NET 2.0 communicating with a robot over Bluetooth radio. The prototype demonstrated an explicit methodology of program authoring, resource management, architecture, and program flow that can be applied to almost any remote interfacing robotics problem. In fact, what is described herein could be applied to an even larger scope of engineering problems.

This research suggests more should be done to study how humans will interface with in-home companion robots by defining the scenario, the environment, and the anticipated interaction based on need. In order to address such a myriad and complex issue, the construction of a software development kit (SDK) is suggested wherein to programmatically define the physicality and constraints necessary to satisfy the required need. The prototype also showed the feasibility of using wireless schemes to communicate intentions to robots at a distance.

Along with speech technologies, this type of application can provide a compliment if speech is compromised, or if an emergency should occur where only a tactile response is possible. It would be of great interest to see if the hardware particular to the mobile phone, computing watch, or other miniature device such as an implant or wearable item could yield more useful and robust results.

The technology demonstrated here and the paradigm suggested in the creation of its processes could very well aid the development of *CompanionAble* and more diverse categories of robots primarily required for interaction with humans who are wholly dependent on them for health, welfare, stewardship, and ultimately survival. A list of some ideas for future research of this idea is:

1. Migrate the code to Windows Embedded or other desired platform such as Android or Linux; in the latter cases, it would be preferred to port the code than use a cross-platform shell,
2. Granularity in the motor motion controls providing economy of programming including time spent and quality of motion,
3. Implement fusion in code NXT sensor clamps creating real-time 3-sensor feedback,
4. Voice activated control and voice response (TTS),
5. Designing and constructing a software development kit (SDK) for programmatic expansion,
6. Including communication over WiFi and location-based services,
7. Integrate the project into a more complex robot.

A final consideration is the level of autonomy a person would like to provide their robot, not only given the task the machine is designed to perform—industry, social, personable, caregiving—but the level of comfort a human would express when witnessing full autonomy. Is it sufficient that a robot, contained to a given area, should remain there and only interact with humans when they cross each other's paths? What about human-to-robotic interaction for recreational, informational, or emergency purposes? Is one satisfied by the safeguards built into the programming based on the understanding of our own cognitive processes? Such concepts have been discussed [6, 9], however, independent of post-priori analysis, what kind of control can the user impose on the robot at runtime? The author argues the user should be allowed to choose what impositions are placed on the robot; otherwise, it will never receive the full acceptance in society that deserves such an ambition.